\documentclass[a4paper,12pt]{article}

\usepackage[font={small}]{caption}
\usepackage{amsmath}
\usepackage{color,graphicx}
\usepackage[top=2.5cm,right=2cm,bottom=2.5cm,left=2cm]{geometry}
\usepackage[colorlinks,linkcolor=blue,citecolor=blue,pagebackref=false]{hyperref}
\usepackage{cite}
\usepackage[parfill]{parskip}
\usepackage{caption}
\usepackage{subfigure}
\usepackage{float}
\usepackage{hyperref}
\usepackage{mathtools}
\usepackage{lineno}
\usepackage{multirow}

\graphicspath{{Figures/}}

\begin{document}

\begin{center}

\textbf{Information, entropy and the paradox of choice:\\A model for understanding human choice behavior}

\vspace{0.5 cm}

Mojtaba Madadi Asl\textsuperscript{1,*}\let\thefootnote\relax\footnotetext{* Corresponding author. Email address: m.madadi@ipm.ir.},
Kamal Hajian\textsuperscript{1,2},
Saeideh Ramezani Akbarabadi\textsuperscript{1},\\
Rouzbeh Torabi\textsuperscript{1,3}, and
Mehdi Sadeghi\textsuperscript{4}
\\
\bigskip
\footnotesize{
\textsuperscript{1}School of Biological Sciences, Institute for Research in Fundamental Sciences (IPM), Tehran, Iran\\
\textsuperscript{2}Department of Physics, Middle East Technical University, Ankara, Turkey\\
\textsuperscript{3}Pegah Dadekavan Sharif Company (TAPSELL), Tehran, Iran\\
\textsuperscript{4}Department of Medical Genetics,\\National Institute of Genetic Engineering and Biotechnology (NIGEB), Tehran, Iran}\\

\vspace{0.7cm}



\end{center}


\begin{center}{\small
\begin{minipage}{14cm}

\begin{center}
\textbf{Abstract}
\end{center}

\vspace{0.2cm}

Choice overload occurs when individuals feel overwhelmed by excessive alternatives during decision making. Although larger choice sets are often assumed to be more satisfying, behavioral evidence reveals an inverted U-shaped relationship between satisfaction and choice set size. However, quantitative frameworks linking information processing to choice satisfaction remain underdeveloped. Here, we develop a simple framework based on relative entropy and effective information to explain this behavior. We propose that satisfaction depends on the probability of finding an ideal option within a choice set and is determined by the informational structure of preferential choice probabilities relative to a baseline state of indifference. Small to moderately sized sets allow efficient comparison and identification of preferred options, thereby maximizing both effective information and satisfaction. As the number of alternatives increases, cognitive limitations increase uncertainty, leading to reduced effective information and satisfaction. This mechanism naturally produces the experimentally observed inverted U-shaped dependence of satisfaction on choice set size. Behavioral experiments across varying choice set sizes closely matched model predictions, suggesting that effective information provides a robust metric for choice satisfaction. These findings offer a principled theoretical account of the paradox of choice and carry broader implications for consumer psychology and human choice behavior.\\


\end{minipage}}
\end{center}



\section{Introduction}

In modern life, individuals encounter an ever-expanding array of choices. From selecting items at a coffee shop or restaurant menu to choosing career paths, planning vacations, purchasing consumer products, or even finding life partners, the multitude of options can be overwhelming~\cite{misuraca2024advantages}. While more choices might intuitively seem advantageous, experimental evidence indicates that an abundance of options is not always beneficial~\cite{reibstein1975number,iyengar2000choice,chernev2003more,schwartz2004doing,schwartz2004paradox,shah2007buying,reutskaja2022choice}. Depending on context~\cite{scheibehenne2010can,greifeneder2010less,chernev2015choice,moser2017no}, larger choice sets often increase cognitive demand and decision time, reducing choice satisfaction~\cite{reutskaja2009satisfaction,diehl2010great,reutskaja2018choice}. This effect can be described as an inverted U-shaped function of choice set size, in which the costs of choice outweigh its benefits. This phenomenon, known as \textit{choice overload} or the \textit{paradox of choice}, is grounded in theories of bounded rationality and psychological choice models~\cite{schwartz2004doing,rieskamp2006extending,pothos2021information,glimcher2022efficiently}. Understanding this effect is critical for designing effective marketing strategies and economic policies aimed at moderating consumer behavior~\cite{chernev2003more,scheibehenne2010can,grant2011too,chernev2015choice,buturak2017choice,moser2017no}.

Experimental evidence has consistently demonstrated that an abundance of choices can paradoxically reduce satisfaction. In a seminal study, for example, consumers shopping in a grocery store reported lower satisfaction with their purchases and were less likely to make a purchase when presented with 24 jam flavors compared with only 6 alternatives~\cite{iyengar2000choice}. Similar choice overload effects have been reported across a wide range of domains, including selections of chocolates~\cite{chernev2003more}, coffee~\cite{mogilner2008mere}, pens~\cite{shah2007buying}, photographs~\cite{gilbert2002decisions,reutskaja2018choice}, and gift boxes~\cite{reutskaja2009satisfaction}, as well as decisions involving pension plans~\cite{huberman2007defined} and prescription drug plans~\cite{hanoch2009much}. These studies indicate that larger assortments often impair decision quality and reduce satisfaction with the chosen option. Moreover, they weaken preference strength, defined as the degree to which individuals favor their selected option relative to the available alternatives. These findings suggest that the subjective benefits of increased variety may be offset by the cognitive costs associated with evaluating a growing number of alternatives.

Several factors may explain why individuals respond negatively to large choice sets, including decision regret, limited cognitive capacity, choice complexity, the need for justification, and time constraints~\cite{scheibehenne2010can}. A widely accepted account is the two-component model, which formalizes the trade-off between the benefits and costs of making a choice~\cite{reutskaja2009satisfaction,grant2011too}. This framework provides a rationale for the inverted U-shaped relationship between satisfaction and the number of alternatives in the decision process~\cite{reutskaja2009satisfaction,grant2011too}. As illustrated in Fig.~\ref{Fig1}A, costs (red) increase more steeply than benefits (blue), yielding choice satisfaction (green) that follows an inverted-U trajectory across set sizes. This relationship can be captured mathematically by a quadratic formulation, reflecting the parabolic nature of satisfaction as a function of choice set size~\cite{moser2017no}.

Despite empirical evidence for choice overload, quantitative frameworks linking information processing to subjective choice satisfaction capable of explaining the generic inverted U-shaped relationship between satisfaction and choice set size remain underdeveloped. To address this gap, in this study we develop an information-theoretic account of choice satisfaction. Motivated by behavioral findings, we propose that satisfaction depends on the probability of finding an ideal option among the available alternatives. Individuals evaluate a choice set according to their preferences, which can be represented as a probability distribution over the available options. When the number of alternatives becomes excessively large, individuals may become effectively indifferent among choices, such that all options are equally likely to be selected. We treat this state as a baseline corresponding to minimal satisfaction (maximal uncertainty). Deviations from this equiprobable state reflect the emergence of informational preferences and, consequently, greater satisfaction.

Within this framework, we define effective information as the difference between these two states, quantifying the extent to which an individual's preferences depart from random choice. By linking choice satisfaction to effective information, the model predicts a non-monotonic dependence of satisfaction on the number of available alternatives. We derive analytical conditions under which this inverted U-shaped relationship emerges and examine how it depends on the informational structure of preferences. To test the model, we conducted a behavioral experiment in which participants made choices from assortments of varying sizes. The experimentally derived choice probabilities closely matched the assumptions of the framework, and the resulting model predictions successfully reproduced the observed pattern of choice satisfaction. Together, these findings establish a quantitative link between information processing and subjective choice satisfaction, demonstrating how cognitive constraints give rise to the paradox of choice and provide a parsimonious, general framework for understanding decision making in environments with abundant alternatives.


\section{Methods}
\subsection{Theoretical model}
\label{theoretical_model}

Our framework assumes that choice satisfaction depends on the amount of information - quantified as effective information, a proxy for satisfaction - that an individual possesses about her preferences within a set of alternatives. Preferences are represented as a probability distribution over available options, with higher probabilities indicating stronger preference for particular items. Under this formulation, satisfaction increases as preferences become more differentiated and decreases as uncertainty in the preference distribution grows, generating a non-monotonic dependence on choice set size. In the following sections, we present the mathematical formulation of the framework.

\begin{figure}[t!]
\centering
\includegraphics[scale = 1.0]{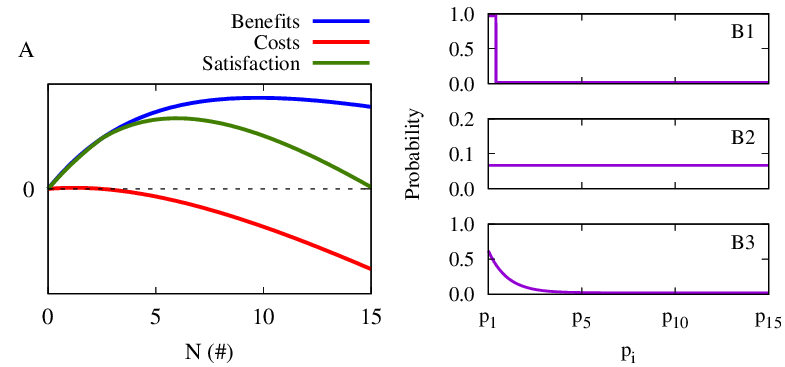}
\caption{{\bf Satisfaction and choice probabilities} (\textbf{A}) Schematic illustration of the two-component model, in which choice satisfaction (green) emerges from a trade-off between the benefits (blue) and costs (red) of choice. (\textbf{B1-B3}) Representative probability distributions that an individual may assign to a choice set of size $N = 15$: a deterministic choice centered on the highest-priority option (B1, $p_1 = 1$), an indifferent choice in which all alternatives are equally likely (B2, $p = \frac{1}{15}$); and an intermediate distribution exhibiting partial preferences for higher-ranked alternatives (B3).} 
\label{Fig1}
\end{figure}


\subsubsection{Ratings and choice probabilities}

Consider an individual confronted with a choice set of size $N$ containing arbitrary items. The individual evaluates features of each item and assigns subjective ratings reflecting her preferences, forming the set $\mathcal{R} = \lbrace r_1, r_2, \dots, r_N \rbrace$, where $r_1$ corresponds to the highest-rated (most preferred) option and $r_N$ to the lowest-rated option. To represent preferences probabilistically, ratings were transformed into normalized choice probabilities using a softmax function~\cite{mckelvey1995quantal,madadiasl2025boltzmann}, $p_i = \exp(\beta r_i) / \sum_{j=1}^{N}\exp(\beta r_j)$ for $i = 1, 2, ..., N$, where $\beta > 0$ is a scaling parameter, controlling the sensitivity of choice probabilities to differences in ratings (throughout this study, we set $\beta = \ln 2$). The resulting set of probabilities is denoted by $\mathcal{P} = \lbrace p_1, p_2, \dots, p_N \rbrace$, where $p_i$ represents the probability of selecting the $i$-th priority and $\sum_{i=1}^{N} p_i = 1$. For analytical tractability, probabilities are ordered by preference rather than item labels, such that $p_1$ corresponds to the most preferred option and $p_N$ to the least preferred.

These probabilities arise from individual comparisons across item features. Heterogeneity among alternatives can introduce uncertainty, leading to deviations from deterministic choice behavior. Prior work has demonstrated how such uncertainty can be quantified using entropy measures derived from observed variability in choice probabilities~\cite{dean2022better,caplin2022rationally}. In the \hyperref[behavioral_experiment]{Behavioral experiment}, participants provided item ratings, which were subsequently transformed into choice probabilities.


\subsubsection{Illustrative examples}

To provide intuition for the proposed framework, we consider three representative scenarios in which an individual chooses one option from a set of $N$ alternatives.
\begin{itemize}
\item \textit{Deterministic choice}: The individual identifies a preferred option with complete certainty and assigns probability $p_1 = 1$ to that option and $p_{-1} = 0$ to all remaining alternatives ($i \neq 1$). The resulting probability distribution is a delta function centered on the preferred option (Fig.~\ref{Fig1}B1). This scenario corresponds to minimal uncertainty and entropy, reflecting maximal information about the individual's preferences. Because the preferred option is unambiguously identified, this case represents maximal choice satisfaction.
\item \textit{Indifferent choice}: The individual is indifferent among all alternatives and assigns equal probability to each option, such that $p_1 = p_2 = ... = p_N = p = \frac{1}{N}$. The resulting uniform distribution (Fig.~\ref{Fig1}B2; $ p = \frac{1}{15}$ for the illustrated example) corresponds to maximal uncertainty and entropy. In this regime, the choice is effectively random and contains minimal information about the individual's preferences.
\item \textit{Partial preference}: The individual exhibits preferences among alternatives but does not choose any option with complete certainty. The resulting probability distribution assigns larger probabilities to more preferred options and smaller probabilities to less preferred alternatives (Fig.~\ref{Fig1}B3). This intermediate regime reflects a finite degree of uncertainty and captures the majority of real-world decisions.
\end{itemize}
These scenarios define different cases of the preference distributions considered in the model. As shown below, their associated entropy values provide a natural basis for quantifying the information contained in an individual's preferences. As a baseline, we consider a state of indifference in which all alternatives are equally likely to be chosen, corresponding to maximum Shannon entropy ($H_{\rm max}$), which represents maximal uncertainty regarding preferences. In contrast, an individual's preference distribution is characterized by an observed entropy ($H_{\rm obs}$), which quantifies the uncertainty associated with her choices. Effective information ($I$) measures the extent to which an individual's preferences deviate from the baseline state of indifference and therefore quantifies the information available in the choice process.


\subsubsection{Shannon entropy}

To quantify the uncertainty inherent in choice probability distributions, we employ the Shannon entropy of a discrete probability distribution, $\mathcal{P} = \lbrace p_1,p_2,...,p_N \rbrace$, which is defined as:
\begin{equation}\label{eq:1}
H(\mathcal{P}) = - \sum_{i = 1}^N p_i \log_2(p_i),
\end{equation}
where $p_i$ denotes the probability of selecting the $i$-th priority. Entropy quantifies the expected uncertainty or information contained in the distribution, measured in bits. A uniform distribution, in which all options are equally probable, achieves maximal entropy, whereas a distribution concentrated on a single option has minimal entropy.


\subsubsection{Effective information and relative entropy}

Effective information ($I$) quantifies the reduction of uncertainty in an observed choice probability distribution relative to a baseline of maximal uncertainty. Let $\mathcal{P} = \lbrace p_1,p_2,...,p_N \rbrace$ denote the \textit{a priori} distribution representing maximal entropy (i.e., equiprobable choices), and $\mathcal{P}^{\prime} = \lbrace p^{\prime}_1,p^{\prime}_2,...,p^{\prime}_N \rbrace$ the \textit{a posteriori} distribution representing the observed choice probabilities. Effective information is defined as the relative entropy of $\mathcal{P}^{\prime}$ with respect to $\mathcal{P}$~\cite{balduzzi2008integrated,madadiasl2025boltzmann}:
\begin{equation}\label{eq:2}
I(\mathcal{P},\mathcal{P}^{\prime}) = H(\mathcal{P}^{\prime} || \mathcal{P}),
\end{equation}
where the relative entropy of $\mathcal{P}^{\prime}$ with respect to $\mathcal{P}$ is defined as follows~\cite{cover1999elements}:
\begin{equation}\label{eq:3}
H(\mathcal{P}^{\prime} || \mathcal{P}) = \sum_i p^{\prime}_i \log_2 \left( \dfrac{p^{\prime}_i}{p_i} \right) .
\end{equation}
Relative entropy is non-negative and equals zero if and only if $\mathcal{P}^{\prime} = \mathcal{P}$. Using Shannon entropy, effective information can equivalently be expressed as~\cite{balduzzi2008integrated}:
\begin{equation}\label{eq:4}
I(\mathcal{P},\mathcal{P}^{\prime}) = H_{\rm max}(\mathcal{P}) - H_{\rm obs}(\mathcal{P}^{\prime}),
\end{equation}
where $H_{\rm max} = \log_2(N)$ corresponds to the maximum entropy of an equiprobable distribution, and $H_{\rm obs}$ is the Shannon entropy of the observed choice distribution given by Eq.~(\ref{eq:1}). We argue that effective information defined in Eq.~(\ref{eq:4}) serves as an appropriate and quantitative metric for assessing choice satisfaction, as it captures the degree to which an individual's preference distribution departs from a baseline state of indifference.


\subsubsection{Optimal choice set size}

To account for cognitive limitations and time constraints in decision making, we assume the existence of an optimal choice set size ($m$) which represents the set size at which an individual can reliably identify her preferred option. In experimental data, $m$ can be estimated from observed choice behavior; in the model, it serves as the turning point for the inverted U-shaped relationship between choice satisfaction and set size. For choice sets smaller than or equal to the optimal size ($N \leq m$), effective information, $I$ in Eq.~(\ref{eq:4}), increases monotonically with set size. That is, for $N < N^\prime \leq m$:
\begin{equation}\label{eq:5}
I(\mathcal{P},N) \leq I(\mathcal{P}^{\prime},N^{\prime}),
\end{equation}
where $\mathcal{P}$ and $\mathcal{P}^\prime$ are the choice probability distributions for set sizes $N$ and $N^\prime$, respectively. For choice sets larger than the optimal size ($N > m$), cognitive constraints prevent complete evaluation of all options. We denote the number of unprocessed or \textit{missed choices} as $s = N - m$, assuming they are assigned equal probabilities due to lack of preference. Consequently, effective information decreases monotonically with set size, such that for $N > N^\prime > m$:
\begin{equation}\label{eq:6}
I(\mathcal{P},N) \leq I(\mathcal{P}^{\prime},N^{\prime}).
\end{equation}
Together, these conditions produce an inverted U-shaped relationship between choice satisfaction and choice set size, with maximal satisfaction occurring near $N^* = m$.


\subsubsection{Model integration}

Computer simulations were performed on a custom code developed in C++11. The simulation code and the Gnuplot command-line scripts used to generate the final versions of all figures are publicly accessible at \url{https://github.com/MMadadiAsl/Choice-satisfaction-model}~\cite{madadiasl2026github}. Running the code enables reproduction of the model predictions presented in the manuscript, including the visualization of the numerical example shown in Figs.~\ref{Fig3} and~\ref{Fig6}, as well as the analytical examples presented in Figs.~\ref{Fig4} and~\ref{Fig5}.


\subsection{Behavioral experiment}
\label{behavioral_experiment}
\subsubsection{Participants}

A total of 240 undergraduate and graduate students (150 females; aged 18-35 years) from multiple universities voluntarily participated in this behavioral experiment. The study was designed to investigate choice overload and satisfaction across six groups (Groups A-F), each of which was assigned a choice set of a specific size ($N = 4, 8, 12, 16, 20,$ and $24$). Participants were evenly distributed across conditions, with $n = 40$ participants assigned to each group. Descriptive statistics of the participants and experimental conditions are provided in Table~\ref{table1}. Prior to participation, all individuals provided written informed consent. The study was conducted in accordance with the Declaration of Helsinki and was approved by the Research Ethics Committee of Tarbiat Modares University (approval ID: IR.MODARES.REC.1405.049). The complete set of experimental materials, raw data, and variable codebook is publicly available at \url{https://doi.org/10.5281/zenodo.20565424}~\cite{madadiasl2026zenodo}. These data were used to generate the experimental results presented in Figs.~\ref{Fig7}-\ref{Fig9}.


\subsubsection{Design and procedure}

\begin{figure}[t!]
\centering
\includegraphics[scale = 0.82]{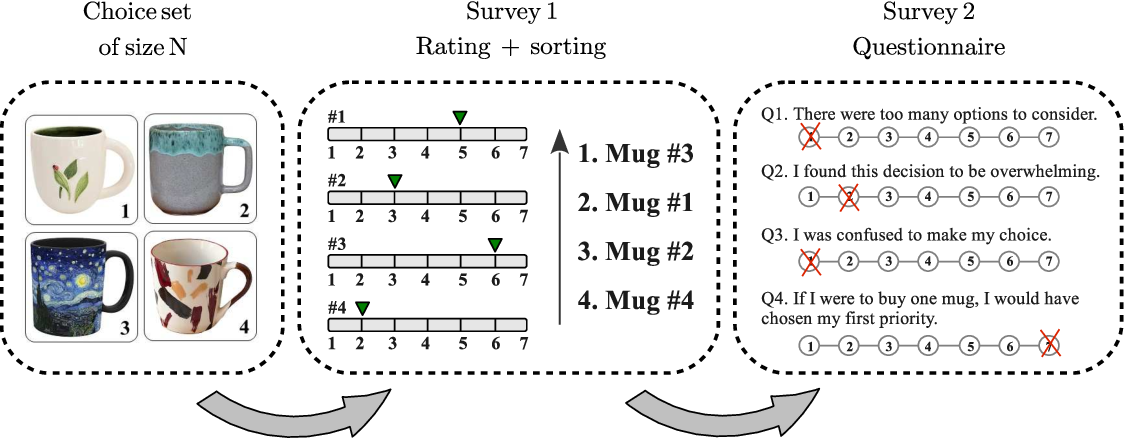}
\caption{\textbf{Experimental design.} Participants were presented with a choice set of size $N$ containing labeled mug images ($N = 4$ example is shown). In Survey 1, they rated the mugs based on interest using a 7-point Likert scale, and then sorted them. In Survey 2, participants answered questions assessing perceived choice overload and satisfaction, also scored on a 7-point Likert scale.}
\label{Fig2}
\end{figure}

Participants completed a decision-making task in which they were presented with choice sets containing different numbers of ceramic mugs (Fig.~\ref{Fig2}, $N = 4$ example) and asked to select the option they most preferred. Choice set sizes varied across groups: Group A ($N = 4$), Group B ($N = 8$), Group C ($N = 12$), Group D ($N = 16$), Group E ($N = 20$), and Group F ($N = 24$). Mugs varied in design, shape, and color but within a specific price range to ensure comparable evaluation difficulty. As illustrated in Fig.~\ref{Fig2}, participants rated and sorted the mugs according to their interest in Survey 1, imagining a real purchase scenario. Ratings were provided on a 7-point Likert scale (1 = strongly dislike, 7 = strongly like) and were subsequently transformed into normalized choice probabilities ($p_i$).

To quantify choice overload and subjective satisfaction, we adapted four questions from prior studies~\cite{agnew2005asset,guo2022can}, administered across all six choice set conditions ($N = 4, 8, 12, 16, 20,$ and $24$) in Survey 2, as demonstrated in Fig.~\ref{Fig2}. Questions included: Q1: ``There were too many options to consider'' (perceived choice abundance), Q2: ``I found this decision to be overwhelming'' (perceived overwhelm), Q3: ``I was confused to make my choice'' (confusion), and Q4: ``If I were to buy one mug, I would have chosen my first priority'' (choice satisfaction). All items were rated on a 7-point Likert scale (1 = strongly disagree, 7 = strongly agree).

\begin{table}[t!]
\caption{{\bf Descriptive statistics of the experiment.} The table summarizes participant grouping, choice set sizes ($N$), age ranges (18-23 (A1), 24-29 (A2), and 30-35 (A3) years), and gender distribution (F: female; M: male) across experimental conditions in which the number of available options varied from $N = 4$ to $N = 24$. Each experimental condition comprised $n = 40$ participants.} 
\centering{
{\normalsize
\begin{tabular}{|c|c|c|c|c|c|c|c|}
\hline
\multirow{2}{*}{Participants} & \multirow{2}{*}{$N$ (\#)} & \multirow{2}{*}{$n$ (\#)} & \multicolumn{3}{|c|}{Age (\%)} & \multicolumn{2}{|c|}{Gender (\%)} \\ \cline{4-8}
 &  & & A1 & A2 & A3 & F & M \\ \hline
Group A & 4 & 40 & 72.5 & 02.5 & 25.0 & 62.5 & 37.5 \\ \hline
Group B & 8 & 40 & 40.0 & 42.5 & 17.5 & 67.5 & 32.5 \\ \hline
Group C & 12 & 40 & 30.0 & 40.0 & 30.0 & 60.0 & 40.0 \\ \hline
Group D & 16 & 40 & 65.0 & 25.0 & 10.0 & 65.0 & 35.0 \\ \hline
Group E & 20 & 40 & 72.5 & 22.5 & 05.0 & 65.0 & 35.0 \\ \hline 
Group F & 24 & 40 & 22.5 & 42.5 & 35.0 & 55.0 & 45.0 \\ \hline 
\end{tabular}}}
\label{table1}
\end{table}	


\subsection{Data analysis}
\label{data_analysis}

All statistical analyses were performed using two-tailed tests with a significance threshold of $\alpha = 0.05$. Questionnaire responses are reported as mean $\pm$ standard deviation (SD) throughout the study. To assess the effect of choice set size on subjective evaluations of the decision-making process, separate one-way analyses of variance (ANOVAs) were conducted for each questionnaire item in Survey 2. Effect sizes were quantified using eta squared ($\eta^2$). Pairwise comparisons were performed using Welch's independent samples $t$-tests between neighboring choice set sizes. Additional comparisons were performed between the smallest and largest choice sets ($N=4$ and $N=24$) for Q1-Q3. For Q4, which was hypothesized to exhibit an inverted-U dependence on choice set size, planned comparisons were conducted between $N=4$ and $N=12$, and between $N=12$ and $N=24$, corresponding to the ascending and descending segments of the predicted relationship.

Model predictions of effective information ($I$) were derived from participants' choice probabilities for each choice set size ($N$). For each set size ($N=4, 8, 12, 16, 20,$ and $24$), the choice probabilities were first averaged across participants, and the effective information was computed using Eq.~(\ref{eq:4}). To characterize the non-monotonic relationship between effective information and choice set size, a second-order polynomial (quadratic) was fitted to the six resulting model predictions according to $I(N) = \beta_0 + \beta_1 N + \beta_2 N^2$, where $\beta_0$, $\beta_1$, $\beta_2$ are coefficients obtained from quadratic fitting. The fitted quadratic provides a descriptive characterization of the inverted-U relationship and allows estimation of the optimal choice set size corresponding to maximal effective information, i.e., $N^* = - \frac{\beta_1}{2 \beta_2}$. Goodness of fit was assessed using the coefficient of determination ($R^2$), which quantifies the proportion of variance in the model predictions explained by the quadratic function.

\section{Results}
\subsection{Model results}
\subsubsection{Effective information as a measure of choice satisfaction}

When choice sets are small, individuals can more readily identify their preferred option and assign probabilities with greater confidence. In the limiting case of a deterministic choice, the probability distribution approaches a delta function centered on the preferred alternative (Fig.~\ref{Fig1}B1), yielding zero observed entropy ($H_{\rm obs} = 0$) and maximal effective information ($I = H_{\rm max}$). This regime corresponds to minimal uncertainty and maximal satisfaction. At the opposite extreme, large choice sets may lead to indifference among alternatives, producing an approximately uniform probability distribution (Fig.~\ref{Fig1}B2). In this case, the observed entropy reaches its maximum value ($H_{\rm obs} = H_{\rm max}$), effective information vanishes ($I = 0$), and choices become effectively random, reflecting maximal uncertainty and minimal satisfaction. Most real-world decisions lie between these extremes and are characterized by probability distributions that assign higher probabilities to preferred alternatives while retaining some degree of uncertainty (Fig.~\ref{Fig1}B3). For such distributions, effective information takes intermediate values and quantifies the extent to which preferences reduce uncertainty relative to a state of indifference. Under our framework, this reduction in uncertainty provides a natural measure of choice satisfaction.

\begin{figure}[h!]
\centering
\includegraphics[scale = 1.0]{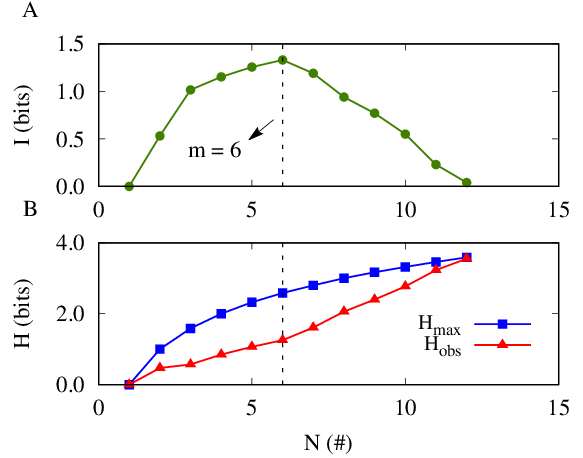}
\caption{{\bf Effective information as a measure of choice satisfaction.} (\textbf{A}) Effective information ($I$) defined in Eq.~(\ref{eq:4}) quantifies choice satisfaction. Using the probability distributions in Table~\ref{table2}, the inverted U-shaped relationship between satisfaction and choice set size ($N$) is reproduced. (\textbf{B}) Maximum entropy ($H_{\rm max}$, blue curve with squares) and observed entropy ($H_{\rm obs}$, red curve with triangles) are plotted for comparison. Vertical dashed lines indicate the optimal choice set size ($m = 6$).}
\label{Fig3}
\end{figure}

In our framework, effective information introduced in Eq.~(\ref{eq:4}) serves as a metric for choice satisfaction. The intuition underlying this relationship is illustrated by the example presented in Table~\ref{table2} and visualized in Fig.~\ref{Fig3}. When only a single option is available ($N = 1$), no meaningful choice can be made and both the maximum and observed entropy are zero ($H_{\rm obs} = H_{\rm max} = 0$), yielding $I = 0$. As the number of available alternatives increases, individuals gain the opportunity to identify options that better match their preferences. Consequently, the observed probability distribution becomes increasingly concentrated around preferred alternatives, resulting in higher effective information and greater satisfaction.

In this illustrative example, the optimal choice set size is fixed at $m = 6$, corresponding to the point at which the individual can most effectively identify and discriminate among alternatives (vertical dashed lines in Fig.~\ref{Fig3}). As shown in Fig.~\ref{Fig3}A, effective information increases monotonically with choice set size for $N \leq m$, consistent with Eq.~(\ref{eq:5}). In this regime, the addition of new alternatives provides useful information and increases the likelihood of identifying a preferred option. Beyond the optimum ($N > m$), however, effective information decreases with choice set size, consistent with Eq.~(\ref{eq:6}). Under our framework, larger sets impose greater cognitive demands, making it increasingly difficult to evaluate all available alternatives and leading to less discriminative probability distributions. Consequently, the information gained from choice declines, producing the characteristic inverted U-shaped relationship between satisfaction and choice set size.

Importantly, we do not assume a cumulative expansion of the choice set. Rather, consistent with previous behavioral studies of choice overload~\cite{iyengar2000choice,chernev2003more,shah2007buying,reutskaja2018choice,reutskaja2022choice}, each choice set size is treated as an independent decision context. Thus, when evaluating a choice set of size $N$, individuals are assumed to have no memory of previously encountered choice sets, their ratings, or the associated choice probabilities. Each decision is therefore characterized by a newly constructed probability distribution reflecting preferences within the current set. Under this assumption, increasing the number of alternatives initially enhances the likelihood of identifying a preferred option, leading to higher effective information. However, beyond an optimal set size, cognitive limitations hinder the evaluation of all available alternatives. As a result, preferences become less discriminative, the observed entropy increases, and effective information declines. This mechanism naturally produces the inverted U-shaped relationship between satisfaction and choice set size that characterizes choice overload.

\begin{table}[t!]
\caption{{\bf Representative choice probabilities across different set sizes.} This table presents example probability distributions ($p_i$) assigned to options in choice sets ranging from $N=1$ to $N=12$, ordered by individual preference. For each set, $p_1$ corresponds to the most preferred option and $p_N$ to the least preferred. The optimal choice set size is fixed at $m = 6$. Maximum entropy ($H_{\rm max}$), observed entropy ($H_{\rm obs}$), and effective information ($I = H_{\rm max}-H_{\rm obs}$) are shown for each set. The table illustrates how effective information increases with choice set size up to the optimal point and decreases for larger sets, reproducing the inverted U-shaped relationship between choice set size and satisfaction (Fig.~\ref{Fig3}A).} 
\centering{
{\scriptsize
\begin{tabular}{|c|c|c|c|c|c|c|c|c|c|c|c|c|c|c|c|}
\hline
$N$ & $p_1$ & $p_2$ & $p_3$ & $p_4$ & $p_5$ & $p_6$ & $p_7$ & $p_8$ & $p_9$ & $p_{10}$ & $p_{11}$ & $p_{12}$ & $H_{\rm max}$ & $H_{\rm obs}$ & $I$ \\ \hline
1 & 1 & \multicolumn{11}{c|}{} & 0.00 & 0.00 & 0.00 \\ \hline
2 & 0.9 & 0.1 & \multicolumn{10}{c|}{} & 1.00 & 0.47 & 0.53 \\ \hline
3 & 0.9 & 0.05 & 0.05 & \multicolumn{9}{c|}{} & 1.58 & 0.56 & 1.02 \\ \hline 
4 & 0.85 & 0.05 & 0.05 & 0.05 & \multicolumn{8}{c|}{} & 2.00 & 0.85 & 1.15 \\ \hline
5 & 0.80 & 0.09 & 0.06 & 0.04 & 0.01 & \multicolumn{7}{c|}{} & 2.32 & 1.06 & 1.26 \\ \hline 
6 & 0.77 & 0.10 & 0.05 & 0.03 & 0.03 & 0.02 & \multicolumn{6}{c|}{} & 2.58 & 1.25 & 1.33 \\ \hline 
7 & 0.70 & 0.10 & 0.05 & 0.05 & 0.04 & 0.03 & 0.03 & \multicolumn{5}{c|}{} & 2.80 & 1.61 & 1.19 \\ \hline 
8 & 0.60 & 0.11 & 0.06 & 0.05 & 0.05 & 0.05 & 0.04 & 0.04 & \multicolumn{4}{c|}{} & 3.00 & 2.06 & 0.94 \\ \hline 
9 & 0.52 & 0.10 & 0.08 & 0.07 & 0.06 & 0.05 & 0.04 & 0.04 & 0.04 &  \multicolumn{3}{c|}{} & 3.17 & 2.40 & 0.77 \\ \hline
10 & 0.43 & 0.09 & 0.08 & 0.07 & 0.07 & 0.06 & 0.05 & 0.05 & 0.05 & 0.05 &  \multicolumn{2}{c|}{} & 3.32 & 2.77 & 0.55 \\ \hline
11 & 0.28 & 0.10 & 0.09 & 0.08 & 0.08 & 0.07 & 0.06 & 0.06 & 0.06 & 0.06 & 0.06 &  \multicolumn{1}{c|}{} & 3.46 & 3.23 & 0.23 \\ \hline
12 & 0.14 & 0.10 & 0.09 & 0.09 & 0.08 & 0.08 & 0.07 & 0.07 & 0.07 & 0.07 & 0.07 & 0.07 & 3.59 & 3.55 & 0.04 \\ \hline
\end{tabular}}}
\label{table2}
\end{table}	

Moreover, the inverted U-shaped pattern should be interpreted as a population-level tendency rather than a deterministic property of every individual decision. A given probability distribution may yield low satisfaction even for relatively small choice sets, whereas highly motivated individuals may successfully evaluate larger sets and maintain high satisfaction. Such variability is expected because individuals differ in decision-making strategies, cognitive resources, and preferences. For example, maximizers tend to search extensively for the best possible option, whereas satisficers often rely on less exhaustive search strategies~\cite{shiner2015maximizers}. Nevertheless, despite these individual differences, behavioral studies consistently report an average inverted U-shaped relationship between satisfaction and choice set size. The example shown in Fig.~\ref{Fig3}A should therefore be viewed as a representative realization of this generic pattern rather than a unique prediction for every decision maker.

Figure~\ref{Fig3}B further illustrates why effective information, rather than observed entropy alone, provides a meaningful measure of satisfaction. Maximum entropy increases monotonically with choice set size according to $H_{\rm max} = \log_2(N)$, reflecting the growing number of possible alternatives. Observed entropy is necessarily bounded by this quantity ($H_{\rm obs} \leq H_{\rm max}$) and therefore cannot, by itself, explain the emergence of an optimal choice set size. Effective information captures the difference between these two quantities and thus quantifies the extent to which an individual's preferences reduce uncertainty relative to a state of indifference. Because both the size of the choice set and the structure of the probability distribution contribute to this quantity, the optimal choice set size is not expected to be universal and may vary across individuals, tasks, and decision-making contexts (also see Fig.~\ref{Fig4}C).


\subsubsection{Optimal set size and missed choices}

As outlined in \hyperref[theoretical_model]{Theoretical model}, we assume the existence of an optimal choice set size ($m$) that reflects the number of alternatives an individual can effectively evaluate. When the number of available options exceeds this limit, some alternatives may receive insufficient attention during the decision-making process. We refer to these alternatives as missed choices and examine how they influence effective information and, consequently, choice satisfaction.

For choice sets smaller than or equal to the optimal size ($N \leq m$), individuals can readily identify their preferred alternative. In the limiting case, the most preferred option is selected with near certainty $p_1 \approx 1$, while all remaining options receive negligible probability (note that this selection process can be described by other forms of probability distributions, as illustrated in Table~\ref{table2} and Fig.~\ref{Fig3}). Under these conditions, the observed entropy approaches zero ($H_{\rm obs} \approx 0$), and Eq.~(\ref{eq:4}) simplifies to:
\begin{equation}\label{eq:7}
I \approx H_{\rm max} = \log_2(N),
\end{equation}
indicating that effective information increases monotonically with choice set size, consistent with Eq.~(\ref{eq:5}). When the number of alternatives exceeds the optimal size ($N > m$), we assume that $s = N - m$ options are insufficiently evaluated (missed choices). The probability assigned to the most preferred option can then be approximated as:
\begin{equation}\label{eq:8}
p_1 \approx 1 - \sum^{m+s}_{i=m+1} p_i,
\end{equation}
where the normalization condition $\sum_{i = 1}^{N} p_i = 1$ holds. For simplicity, we assume that the relative probabilities assigned to the $m$ effectively processed options remain unchanged, whereas the additional $s$ missed choices receive equal probabilities because the individual lacks sufficient information to discriminate among them. Accordingly, one can write:
\begin{equation}\label{eq:9}
p_i = \frac{K(s)}{s}, \qquad i \in\{m+1,m+2,\dots , m+s\},
\end{equation}
where $K(s)$ quantifies the total probability mass assigned to the missed choices and satisfies $0 \leq K(s) \leq 1$. Intuitively, small values of $K(s)$ indicate that missed alternatives contribute little to the decision process, whereas larger values imply that these alternatives increasingly influence the allocation of choice probabilities. To capture this behavior, we consider the following form:
\begin{equation}\label{eq:10}
K(s) = \left(\frac{s}{s+a} \right) ^ b,
\end{equation}
where $a > 0$ and $b \geq 1$ are free parameters. This formulation ensures that the collective influence of missed choices increases with their number while preserving equiprobability among them.

\begin{figure}[t!]
\centering
\includegraphics[scale = 1.0]{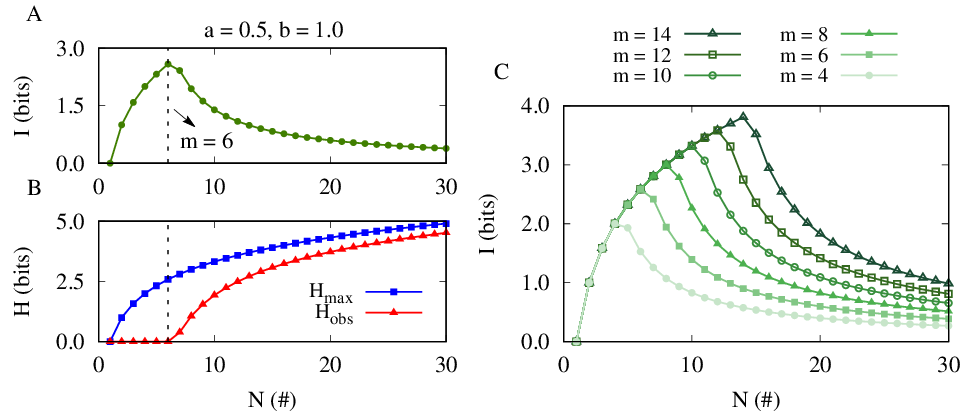}
\caption{{\bf Optimal set size and missed choices determine satisfaction.} (\textbf{A}) Effective information as a function of choice set size using the probability distribution defined in Eq.~(\ref{eq:8}) with $a = 0.5$ and $b = 1.0$, illustrating the influence of optimal set size and missed choices on the inverted U-shaped satisfaction curve. (\textbf{B}) Maximum (blue curve with squares) and observed (red curve with triangles) entropies corresponding to the calculated probabilities. Vertical dashed lines indicate the optimal choice set size ($m = 6$). (\textbf{C}) Effective information for different values of optimal choice set size, demonstrating how the peak of satisfaction shifts with $m$, for example, across individuals, tasks, and decision-making contexts.}
\label{Fig4}
\end{figure}

Figure~\ref{Fig4} illustrates the resulting behavior for $m = 6$, $a = 0.5$ and $b = 1.0$. Effective information initially increases with choice set size, reaches a maximum at the optimal size, and subsequently declines as additional alternatives contribute to the pool of missed choices (Fig.~\ref{Fig4}A). The corresponding maximum and observed entropies are shown in Fig.~\ref{Fig4}B. Importantly, the inverted U-shaped relationship is robust to variations in the optimal choice set size, which primarily shift the location of the peak without altering the overall qualitative behavior (Fig.~\ref{Fig4}C).

\begin{figure}[t!]
\centering
\includegraphics[scale = 1.0]{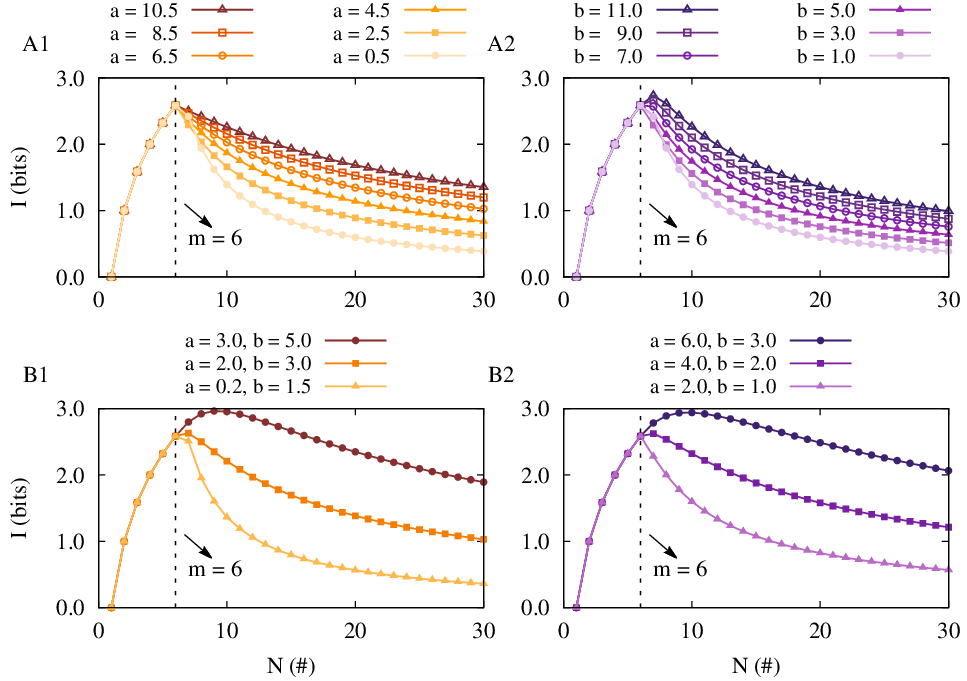}
\caption{{\bf Missed choices parameters influence satisfaction behavior.} (\textbf{A1}) Variation of parameter $a$ in Eq.~(\ref{eq:10}) affects the probability assigned to missed choices, thereby modulating the inverted U-shaped relationship between satisfaction and choice set size for $m = 6$, and $b = 1.0$.  (\textbf{A2}) Variation of parameter $b$ in Eq.~(\ref{eq:10}) with fixed $a = 0.5$ illustrates a similar effect on the shape of the satisfaction curve. (\textbf{B1}) Combined variation of $a$ and $b$ in Eq.~(\ref{eq:10}) with $a < b$ produces different forms of the inverted U-shaped satisfaction curve for $m=6$. (\textbf{B2}) Variation with $a > b$ shows how the relative magnitude of missed choice parameters modulates the peak and asymmetry of the satisfaction function. Vertical dashed lines denote the optimal choice set size ($m = 6$).}
\label{Fig5}
\end{figure}

The treatment of missed choices plays a central role in shaping the satisfaction curve. As shown in Fig.~\ref{Fig5}A1 and A2, varying the parameters $a$ and $b$ changes how rapidly probability mass is transferred to missed alternatives, thereby modulating the decline in effective information beyond the optimal choice set size. Larger values of either parameter generally produce a more gradual decrease in satisfaction as the number of alternatives increases. Finally, Fig.~\ref{Fig5}B1 and B2 demonstrates that different combinations of $a$ and $b$ generate a family of inverted U-shaped relationships. Although the precise location, width, and asymmetry of the peak depend on the parameter values, the qualitative prediction of the model remains unchanged: satisfaction initially increases as additional alternatives provide useful information, but decreases once the number of options exceeds the individual's effective processing capacity.

\begin{figure}[t!]
\centering
\includegraphics[scale = 1.0]{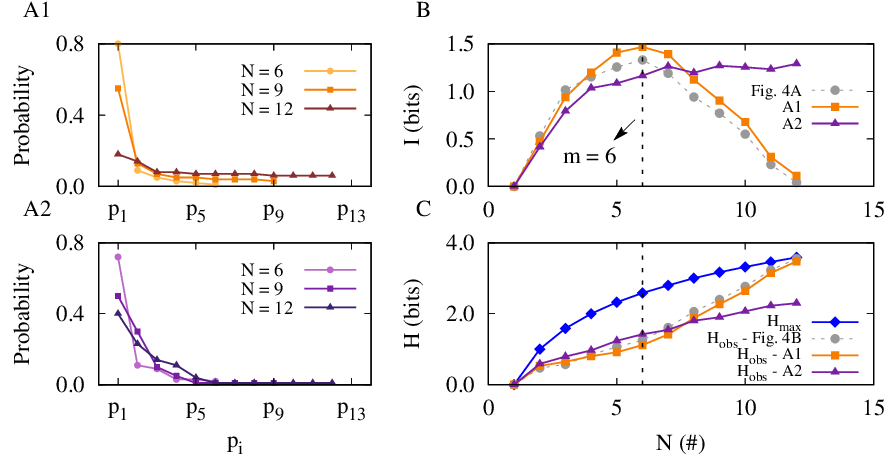}
\caption{{\bf Informational structure of choice probabilities in the model.} (\textbf{A1, A2}) Representative choice probabilities ($p_i$) for three choice set sizes ($N =$ 6, 9, and 12). A distribution that asymptotically approaches equal choice probabilities for all items as the size of the choice set increases (A1) reproduces the inverted U-shaped relationship between satisfaction and choice set size (see panel B, orange curve with squares), whereas a non-uniformly decreasing distribution that maintains relatively high probabilities for certain items and low probabilities for others (A2) fails to reproduce this pattern (panel B, violet curve with triangles). (\textbf{B}) Effective information Eq.~(\ref{eq:4}) quantifies choice satisfaction using the distributions from A1 (orange curve with squares), A2 (violet curve with triangles), and Table~\ref{table2} (gray dashed curve; reproduced from Fig.~\ref{Fig3}A) for comparison. (\textbf{C}) Corresponding maximum entropy (blue curve with diamonds) and observed entropies (orange curve with squares for A1, violet curve with triangles for A2, gray dashed curve from Table~\ref{table2}) illustrate how the structure of the probability distribution affects satisfaction. Vertical dashed lines indicate the optimal choice set size ($m = 6$ in this example).}
\label{Fig6}
\end{figure}


\subsubsection{Informational structure of choice probabilities in the model}

Our model predicts that effective information ($I$) described in Eq.~(\ref{eq:4}) captures the inverted U-shaped relationship between satisfaction and choice set size. Figure~\ref{Fig6} illustrates how this pattern arises from the underlying structure of choice probabilities. Maximum entropy ($H_{\rm max}$) naturally increases with the number of alternatives. In contrast, observed entropy ($H_{\rm obs}$) depends critically on the distribution of choice probabilities. As shown in Fig.~\ref{Fig6}A1, a distribution that converges asymptotically toward equal probabilities for all items across increasingly large choice sets reproduces the inverted U-shaped relationship for satisfaction (Fig.~\ref{Fig6}B, orange curve with squares). This pattern closely aligns with the choice probabilities from Table~\ref{table2} and Fig.~\ref{Fig3}B (gray dashed curve in Fig.~\ref{Fig6}B). The corresponding observed entropy increases with choice set size (Fig.~\ref{Fig6}C, orange curve with squares), similar to the trend in Fig.~\ref{Fig3}B (gray dashed curve in Fig.~\ref{Fig6}C). By contrast, a non-uniformly decreasing distribution that continues to assign high probabilities to a subset of items and low probabilities to missed choices (Fig.~\ref{Fig6}A2) fails to reproduce the inverted U-shaped satisfaction curve (Fig.~\ref{Fig6}B, violet curve with triangles) due to the saturation of observed entropy (Fig.~\ref{Fig6}C, violet curve with triangles). These results demonstrate that the informational structural of choice probabilities is a key determinant of how satisfaction emerges as the number of alternatives increases.


\subsection{Experimental results}

\begin{figure}[t!]
\centering
\includegraphics[scale = 1.0]{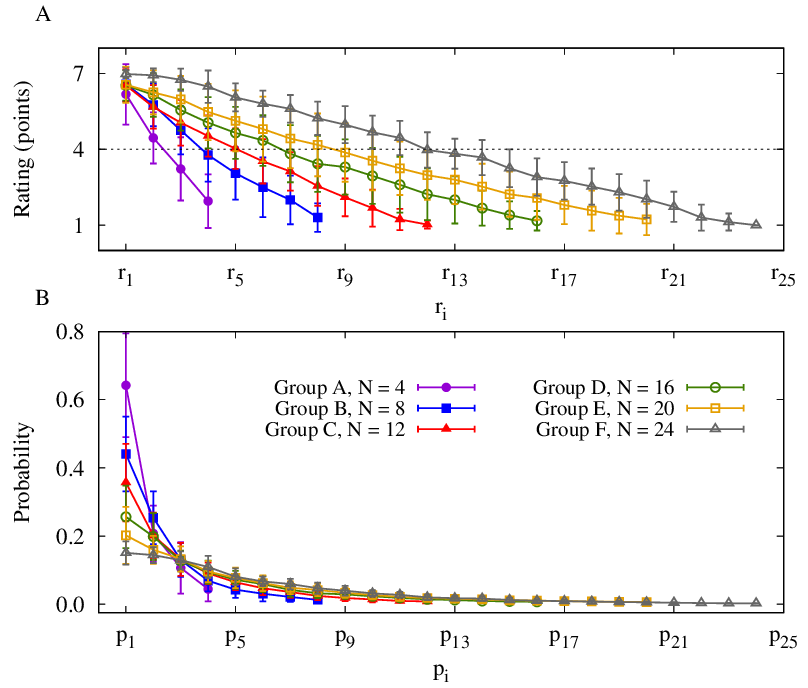}
\caption{{\bf Preference ratings and derived choice probabilities across choice set sizes.} (\textbf{A}) Average sorted ratings ($r_i$) assigned to mugs by participants in Groups A-F, corresponding to choice set sizes ranging from $N = 4$ to $N = 24$. Ratings were provided on a 7-point Likert scale, indicating participants' preference for each mug (1 = strongly dislike, 7 = strongly like), with dashed horizontal line indicating the midpoint of the scale (score = 4). Sorting was performed independently within each participant according to preference rank before averaging across participants. (\textbf{B}) Choice probabilities ($p_i$) derived from the ratings shown in panel A. These probability distributions were subsequently used to compute effective information. Error bars denote standard deviations across participants ($n = 40$ per condition).} 
\label{Fig7}
\end{figure}

To validate our model predictions, we conducted a behavioral experiment investigating choice overload and satisfaction, as described in \hyperref[behavioral_experiment]{Behavioral experiment}. Participants completed a decision-making task in which they were presented with sets of mug images and asked to select the option they most preferred. Choice set sizes varied across groups: Group A ($N=4$), Group B ($N=8$), Group C ($N=12$), Group D ($N=16$), Group E ($N=20$), and Group F ($N=24$). In Survey 1, participants rated each mug according to their level of interest, as if considering a purchase, using a 7-point Likert scale (1 = strongly dislike, 7 = strongly like). The average ratings ($r_i$) for all six groups are shown in Fig.~\ref{Fig7}A. These ratings were then transformed into normalized choice probabilities ($p_i$), as described in \hyperref[theoretical_model]{Theoretical model}, presented in Fig.~\ref{Fig7}B. The structure of these calculated probabilities closely resembles the representative probability distributions shown in Fig.~\ref{Fig6}A1, providing strong empirical support for the underlying assumptions of our model.

In Survey 2, participants completed four questionnaire items assessing perceived choice overload and choice satisfaction, where responses were recorded on a 7-point Likert scale (1 = strongly disagree, 7 = strongly agree). The questionnaire responses revealed significant effects of choice set size on all four dimensions of the decision-making experience. One-way ANOVA demonstrated significant main effects of choice set size for perceived abundance of options (Q1: $F(5,234) = 28.49$, $p < 0.001$, $\eta^2 = 0.378$), perceived overwhelm (Q2: $F(5,234) = 8.39$, $p < 0.001$, $\eta^2 = 0.152$), confusion (Q3: $F(5,234) = 15.65$, $p < 0.001$, $\eta^2 = 0.251$), and choice satisfaction (Q4: $F(5,234) = 7.13$, $p < 0.001$, $\eta^2 = 0.132$). These findings indicate that the number of available alternatives significantly influenced both the cognitive and affective components of the decision-making process.

\begin{figure}[t!]
\centering
\includegraphics[scale = 1.0]{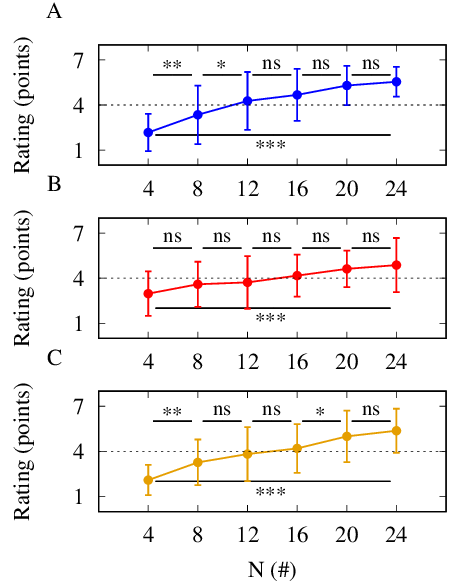}
\caption{\textbf{Behavioral measures of choice overload increase with choice set size.} Mean responses ($\pm$ SD) to Survey 2 items assessing perceived abundance of options (\textbf{A}; Q1: ``There were too many options to consider''), perceived overwhelm (\textbf{B}; Q2: ``I found this decision to be overwhelming''), and confusion (\textbf{C}; Q3: ``I was confused to make my choice'') across six choice set size conditions ($N=4, 8, 12, 16, 20,$ and $24$). Responses were provided on a 7-point Likert scale (1 = strongly disagree, 7 = strongly agree). Larger assortments were associated with progressively greater perceptions of overload, overwhelm, and confusion. One-way ANOVA revealed significant effects of choice set size for all three measures (A: $F(5,234)=28.49$, $p<0.001$; B: $F(5,234)=8.39$, $p<0.001$; C: $F(5,234)=15.65$, $p<0.001$). Error bars indicate standard deviations across participants ($n = 40$ per condition). Horizontal bars indicate pairwise comparisons between adjacent choice set sizes and between the smallest ($N = 4$) and largest ($N = 24$) choice sets. Asterisks denote statistical significance, i.e., $^{***}p < 0.001$, $^{**}p < 0.01$,	$^*p < 0.05$, and ns: not significant.} 
\label{Fig8}
\end{figure}

For perceived abundance of options (Q1), as shown in Fig.~\ref{Fig8}A, ratings increased monotonically with choice set size, indicating that participants increasingly felt that there were too many alternatives to consider as the assortment expanded. Pairwise comparisons revealed significant increases between $N = 4$ and $N = 8$ ($t = -3.23$, $p < 0.01$) and between $N = 8$ and $N = 12$ ($t = -2.14$, $p < 0.05$), whereas subsequent neighboring comparisons did not reach significance. Nevertheless, the difference between the smallest ($N = 4$) and largest ($N = 24$) choice sets was highly significant ($t = -13.49$, $p < 0.001$), demonstrating a substantial cumulative increase in perceived overload. A similar pattern was observed for perceived overwhelm (Q2), as demonstrated in Fig.~\ref{Fig8}B. Although none of the neighboring comparisons reached statistical significance, the overall ANOVA was significant and the difference between the smallest and largest choice sets was highly significant ($t = -5.16$, $p < 0.001$). These results suggest that the subjective burden associated with the decision process accumulates progressively as more alternatives become available. Choice confusion (Q3) likewise increased with choice set size (Fig.~\ref{Fig8}C). Participants reported significantly greater confusion at $N = 8$ than at $N = 4$ ($t = -4.08$, $p < 0.001$), and confusion further increased between $N = 16$ and $N = 20$ ($t = -2.15$, $p < 0.05$). The contrast between the smallest and largest choice sets was again highly significant ($t = -11.66$, $p < 0.001$).

\begin{figure}[t!]
\centering
\includegraphics[scale = 1.0]{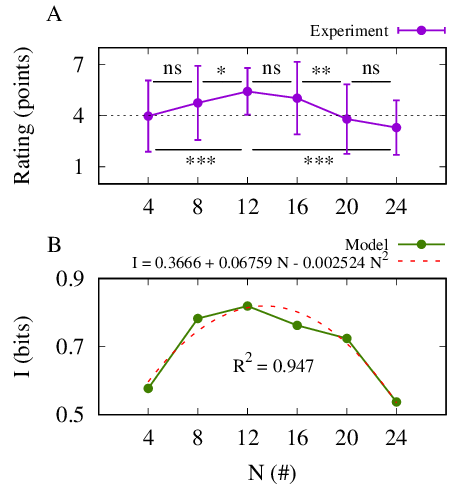}
\caption{\textbf{Behavioral and model characterization of choice satisfaction.} (\textbf{A}) Mean choice satisfaction ratings ($\pm$ SD) across six choice set size conditions ($N=4, 8, 12, 16, 20,$ and $24$). Satisfaction was assessed using the statement ``If I were to buy one mug, I would have chosen my first priority'' in Survey 2 (Q4). Responses were provided on a 7-point Likert scale (1 = strongly disagree, 7 = strongly agree). Satisfaction exhibited a significant dependence on choice set size (one-way ANOVA: $F(5,234)=7.13$, $p<0.001$, $\eta^2=0.132$), increasing from small to intermediate assortments and declining for larger sets, thereby producing an inverted U-shaped relationship. Pairwise differences are indicated by horizontal bars ($^{***}p < 0.001$, $^{**}p < 0.01$,	$^*p < 0.05$, and ns: not significant). Error bars denote standard deviations across participants ($n = 40$ per condition). (\textbf{B}) Model predictions obtained from the effective information ($I$) framework introduced in Eq.~(\ref{eq:4}) using experimentally measured choice probabilities presented in Fig.~\ref{Fig7}B. Each point represents $I$ computed for a given choice set size (green solid curve). The model reproduces the inverted U-shaped relationship observed experimentally in A, with effective information peaking at $N = 12$. The dashed red curve shows a quadratic fit to the model outputs, which provides an excellent description of the model predictions $R^2 = 0.947$ and yields a maximum at $N^* = 13.39$. The close correspondence between panels A and B suggests that effective information captures key features of subjective choice satisfaction across varying choice set sizes.}
\label{Fig9}
\end{figure}

Together, the results of Q1-Q3 indicate that increasing choice set size systematically elevates the cognitive demands of decision making. As the number of alternatives increased, participants were more likely to perceive the choice environment as containing too many options, to report feeling overwhelmed by the decision, and to experience greater confusion when selecting among alternatives. These findings provide direct behavioral evidence for the emergence of choice overload in larger assortments. These results are consistent with the model proposed here, in which larger assortments increase uncertainty and reduce the effective information available for discriminating among alternatives. As a consequence, the cognitive costs associated with evaluating numerous alternatives eventually outweigh the benefits of having more options available, contributing to the decline in choice satisfaction observed for large choice sets.

Accordingly, choice satisfaction (Q4) exhibited a non-monotonic relationship with choice set size, as shown in Fig.~\ref{Fig9}A. Satisfaction increased from smaller to intermediate-sized assortments and subsequently declined as the number of alternatives continued to increase. Pairwise comparisons revealed a significant increase in satisfaction between $N = 8$ and $N = 12$ ($t = -2.20$, $p < 0.05$), followed by a significant decrease between $N = 16$ and $N = 20$ ($t = 2.63$, $p < 0.01$). Consistent with the predicted inverted-U relationship, additional planned comparisons showed that satisfaction was significantly higher at $N = 12$ than at both $N = 4$ ($t = -3.66$, $p < 0.001$) and $N = 24$ ($t = 6.36$, $p < 0.001$). These findings identify an intermediate choice set size ($N = 12$) at which subjective satisfaction is maximized.

Taken together, the behavioral results support the central prediction of the proposed model. Increasing the number of available alternatives provides greater opportunity to identify a preferred option, which contributes positively to satisfaction at small and intermediate choice set sizes. Simultaneously, however, larger assortments increase perceived overload, overwhelm, and confusion, as evidenced by the monotonic increases observed in Fig.~\ref{Fig8}A-C. Satisfaction therefore reflects a trade-off between the benefits of increased opportunity and the cognitive costs associated with evaluating a growing number of alternatives. At small choice set sizes, satisfaction is constrained by limited opportunity to find an ideal option, whereas at large choice set sizes it is reduced by increasing cognitive burden. The resulting balance gives rise to the inverted-U relationship observed in Fig.~\ref{Fig9}A and provides empirical support for the theoretical account of the paradox of choice developed in this study.

The model predictions qualitatively reproduced the experimentally observed inverted-U relationship between choice set size and satisfaction (cf. Fig.~\ref{Fig9} A and B). Fitting a quadratic function (see \hyperref[data_analysis]{Data analysis}) to the six model outputs (Fig.~\ref{Fig9}B, dashed red curve) corresponding to the six set sizes $N = 4, 8, 12, 16, 20,$ and $24$, yielded the coefficients $\beta_0 = 0.3666$, $\beta_1 = 0.06759$, and $\beta_2 = - 0.002524$, with a coefficient of determination $R^2 = 0.947$. This indicates that the quadratic function accounts for the majority of variance in the model predictions. The peak of the fitted curve occurred at approximately $N^* = 13.39$, closely corresponding to the region where participants reported maximal choice satisfaction ($N = 12$), representing optimal choice set size. This analysis confirms that the model captures the non-monotonic dependence of effective information on choice set size and provides a mechanistic explanation for the inverted-U pattern observed in the behavioral data.


\section{Discussion}

In modern life, individuals are frequently confronted with an abundance of alternatives, yet greater choice does not necessarily translate into greater satisfaction. Consistent with the well-established phenomenon of choice overload~\cite{reibstein1975number,iyengar2000choice,chernev2003more,schwartz2004doing,schwartz2004paradox,shah2007buying,reutskaja2022choice}, our behavioral results revealed that choice satisfaction followed an inverted U-shaped relationship with choice set size, increasing from small to intermediate assortments before declining as the number of alternatives became large. At the same time, larger assortments elicited progressively stronger perceptions of overload, overwhelm, and confusion, indicating that the cognitive demands of evaluating multiple alternatives increase systematically with choice set size. To account for these observations, we developed an information-theoretic framework in which satisfaction is linked to the effective information contained in an individual's preference distribution relative to a baseline state of indifference. The model successfully reproduced the characteristic inverted U-shaped dependence of satisfaction on assortment size and closely matched the behavioral data. Our approach differs - in both scope and objective - from previous entropy-based studies that focused on collective dynamics among interacting decision makers~\cite{gupta2024entropy}. Our model provides a quantitative account of satisfaction at the level of the individual, directly relating subjective choice experiences to the informational structure of preferences. Together, these findings suggest that choice satisfaction emerges from a balance between the informational benefits of having more alternatives and the cognitive costs associated with processing them.

Satisfaction is expected to be low at the two extremes of choice. When only a single option is available, individuals have little meaningful freedom to choose. Conversely, when faced with an overwhelming number of alternatives, cognitive limitations may prevent effective evaluation, making choices increasingly arbitrary. In both cases, the information available for distinguishing among options is limited. Satisfaction therefore emerges from the ability to compare alternatives and identify a preferred option. Consistent with this view, our framework predicts that satisfaction is maximized at an intermediate choice set size. Beyond this point, the burden of evaluating more options reduces satisfaction, giving rise to the characteristic inverted U-shaped relationship between satisfaction and choice set size. The present framework was developed to explain this generic pattern independently of any specific product category or decision context. In the choice overload literature, satisfaction is often distinguished into \textit{outcome satisfaction}, referring to the quality of the final choice, and \textit{process satisfaction}, referring to the experience of choosing itself~\cite{reutskaja2009satisfaction}. Our model is most closely related to process satisfaction, as it quantifies the information available during decision making and the extent to which preferences can be differentiated from a state of indifference. However, because process and outcome satisfaction are positively correlated~\cite{reutskaja2009satisfaction}, the principles underlying our framework may also provide insights into broader evaluations of decision quality.

Within our information-theoretic framework, maximum entropy corresponds to a state of maximum uncertainty, where all alternatives are treated as equally likely and no option is preferred over another. As information about available alternatives increases, uncertainty is reduced and preferences become more differentiated, thereby increasing the ability to make informed choices and enhancing satisfaction. The paradox of choice emerges because this relationship is not monotonic: adding alternatives initially increases the informational value of choice by improving the opportunity to identify a preferred option, but beyond a certain point the cognitive demands of evaluating additional alternatives begin to outweigh these benefits. In our model, this transition is captured by the assumption that, once the number of alternatives exceeds an individual's processing capacity, unprocessed or missed options are increasingly treated as equally likely, causing the choice probability distribution to become more uniform and effective information to decline. This assumption is supported by the findings of our behavioral experiment as well as evidence that limitations in attention, working memory, and information-processing capacity constrain decision making in large assortments~\cite{payne1993adaptive,rieskamp2006extending,reutskaja2018choice,hu2024choice}. The resulting increase in entropy provides a conceptual link to random utility models, such as the multinomial logit framework, in which choices reflect both deterministic preferences and random influences~\cite{hausman1984specification,mcfadden2001economic}. As the ability to discriminate meaningful utility differences diminishes, choices become increasingly driven by the random component, approaching a state of maximum entropy. Although the assumption of increasing indifference successfully reproduces the generic inverted U-shaped relationship between choice set size and satisfaction, it should not be regarded as universally applicable. Decision heuristics, default effects, salience, prior knowledge, brand familiarity, and contextual influences can all produce non-uniform choice distributions even in large assortments~\cite{milosavljevic2012relative,bordalo2013salience,nedungadi1990recall,hadar2014knowledge,korhonen2018context,misuraca2019role,misuraca2021more}. Future work could therefore extend the framework by incorporating alternative probability structures and examining the conditions under which the equal-probability assumption holds or breaks down.

Reducing uncertainty about alternatives requires evaluating their attributes, a process constrained by cognitive capacity and time~\cite{haynes2009testing,hu2024choice}. Both the number and complexity of options increase the cognitive demands of decision making, requiring more attention, memory, and comparison effort. When these demands exceed an individual's processing capacity, preferences become less differentiated, choices appear more random, and effective information declines, reflected in higher observed entropy. This mechanism explains why satisfaction initially rises with more options - additional alternatives provide opportunities to identify a preferred choice - but eventually declines when cognitive demands outweigh benefits. The optimal choice set size ($m$) is therefore not universal, varying with individual traits such as expertise, cognitive capacity, and decision strategy, as well as contextual factors like time pressure, stakes, and choice complexity. Nevertheless, when averaged across individuals, these heterogeneous constraints generate the robust inverted U-shaped relationship between satisfaction and choice set size observed across domains ranging from consumer products including chocolates~\cite{chernev2003more}, coffee~\cite{mogilner2008mere}, pens~\cite{shah2007buying}, photographs~\cite{gilbert2002decisions,reutskaja2018choice}, gift boxes~\cite{reutskaja2009satisfaction} to consequential decisions such as romantic partners~\cite{d2017there,thomas2022agony,apostolou2024mate}, pension plans~\cite{huberman2007defined}, prescription drugs~\cite{hanoch2009much}, charitable giving~\cite{scheibehenne2009moderates}, and recommendations~\cite{bollen2010understanding}. Viewed through the lens of effective information, overwhelming alternatives diminishes the informational value of choice itself, reducing satisfaction despite increased freedom. Extensions of the framework could allow $m$ to emerge endogenously using rational inattention principles~\cite{sims2003implications,matvejka2015rational} and multinomial logit models~\cite{hausman1984specification,mcfadden2001economic}, enhancing behavioral realism. In addition, decision making may differ between physical and online environments~\cite{moser2017no}. Online contexts with external cues such as ratings and reviews may alter the location of the optimal set size by reducing search costs, providing an important avenue for future research on how environmental factors influence choice satisfaction and overload.

Despite its grounding in information theory, the present framework has several limitations that warrant further investigation. First, although the model reproduces the inverted U-shaped relationship between satisfaction and choice set size, it treats the optimal choice set size as a fixed parameter and does not explicitly model how it may dynamically change as the number of alternatives, cognitive demands, or available time vary. Second, the model assumes a simplified representation of alternatives and does not account for multi-dimensional item attributes. In practice, products often differ across numerous features, and the complexity of these attributes may substantially influence information processing and satisfaction. Third, our framework considers a generalized decision-making context and does not distinguish between physical and online choice environments. In online settings, external information such as ratings, reviews, and recommendation systems may reduce uncertainty and alter the relationship between choice set size and satisfaction~\cite{moser2017no}. Incorporating these factors into future extensions of the model may improve its applicability to real-world decision-making scenarios and provide a more comprehensive account of choice overload.

From a psychological perspective, consumers often simplify complex decisions by first forming a \textit{consideration set} - a smaller subset of options that are seriously evaluated before a final choice is made~\cite{payne1976task,nedungadi1990recall,kardes2002consideration,goodman2013help,hadar2014knowledge}. At first glance, this process may appear inconsistent with our assumption that individuals assign probabilities across an entire choice set. However, the present framework can be readily adapted to accommodate consideration set formation. Rather than applying the model to all available alternatives, it could be applied to a reduced set generated through simple heuristics, such as price, brand familiarity, or perceived relevance~\cite{alvarez2014choice,misuraca2019role,misuraca2021more}. Satisfaction would then be determined by the information contained within the consideration set rather than the full assortment. This extension would not alter the central prediction of an inverted U-shaped behavior of satisfaction. Instead, it would provide a more psychologically realistic account of how individuals cope with large assortments and suggests that the formation of a consideration set may itself contribute to the cognitive costs underlying choice overload.

Our model could be further strengthened by accounting for heterogeneity in decision-making strategies. Although the framework captures the average inverted U-shaped relationship between satisfaction and choice set size, individuals differ substantially in how they approach decisions. For example, maximizers tend to invest greater effort in searching for the best possible option, whereas satisficers are more likely to settle for an option that is merely acceptable~\cite{shiner2015maximizers}. Such differences may shift the optimal choice set size and alter the point at which choice overload emerges. In addition, contextual factors, such as how options are organized and presented, can influence cognitive load and decision quality. Categorization and structured presentation of alternatives may facilitate information processing and reduce overload in large assortments~\cite{mogilner2008mere}. Incorporating these individual and contextual differences into future versions of the model may provide a more comprehensive account of choice satisfaction and its variability across decision environments.

Ultimately, despite its simplicity, our model can account for a key phenomenon in consumer decision making: the inverted U-shaped relationship between satisfaction and choice set size. This is consistent with evidence showing that excessive choice can reduce purchase satisfaction and increase decision regret~\cite{sarver2008anticipating,buturak2017choice}. Importantly, the model provides a quantitative and conceptually transparent account of why \textit{more choice is not always better}. Beyond its theoretical value, this perspective has practical implications for the design of decision environments. Rather than maximizing product variety, retailers and marketers may benefit from optimizing assortment size and structuring information in ways that facilitate effective comparison among alternatives~\cite{sharma2017switching,mogilner2008mere}. More broadly, our results suggest that choice satisfaction depends not only on the number of available options but also on the extent to which individuals can meaningfully process the information those options provide.


\section*{CRediT Author Statement}

\textbf{Mojtaba Madadi Asl:} Methodology, Formal analysis, Investigation, Visualization, Writing - original draft, Writing - review \& editing, Project administration. \textbf{Kamal Hajian:} Methodology, Formal analysis, Investigation, Writing - original draft, Writing - review \& editing. \textbf{Saeideh Ramezani Akbarabadi:} Methodology, Formal analysis, Data Curation, Writing - review \& editing. \textbf{Rouzbeh Torabi:} Methodology, Formal analysis, Writing - review \& editing. \textbf{Mehdi Sadeghi:} Conceptualization, Methodology, Formal analysis, Writing - original draft, Writing - review \& editing, Supervision.


\section*{Declaration of Competing Interests}

The authors declare that the research was conducted in the absence of any commercial or financial relationships that could be construed as a potential conflict of interest.


%


\section*{Ethics Statement}

All participants provided written informed consent prior to participation in the behavioral experiment. The study was conducted in accordance with the Declaration of Helsinki and was approved by the Research Ethics Committee of Tarbiat Modares University (approval ID: IR.MODARES.REC.1405.049).


\section*{Data Availability}

All data used to produce the model predictions were generated via numerical simulations. The simulation code and the Gnuplot command-line scripts used to generate the final versions of all figures in the manuscript are publicly accessible at \url{https://github.com/MMadadiAsl/Choice-satisfaction-model}~\cite{madadiasl2026github}. All experimental data, including the raw dataset, study materials, and variable codebook are publicly available at \url{https://doi.org/10.5281/zenodo.20565424}~\cite{madadiasl2026zenodo}.


{\footnotesize\bibliography{references}}
\bibliographystyle{vancouver}
\addcontentsline{toc}{section}{References}

\end{document}